\documentclass[11pt]{article}

\usepackage[utf8]{inputenc}
\usepackage[T1]{fontenc}
\usepackage{amsmath,amssymb,amsthm}
\usepackage{mathtools}
\usepackage{bm}
\usepackage{geometry}
\usepackage{natbib}
\usepackage{graphicx}
\usepackage{booktabs}
\usepackage{enumitem}
\usepackage{algorithm}
\usepackage{algorithmic}
\usepackage{kpfonts}
\usepackage{hyperref}
\usepackage{cleveref}

\newtheorem{theorem}{Theorem}[section]
\newtheorem{proposition}[theorem]{Proposition}

\newtheorem{corollary}[theorem]{Corollary}
\newtheorem{definition}[theorem]{Definition}
\newtheorem{assumption}[theorem]{Assumption}
\newtheorem{remark}[theorem]{Remark}

\newcommand{\R}{\mathbb{R}}

\newcommand{\E}{\mathbb{E}}
\newcommand{\Var}{\mathrm{Var}}
\newcommand{\Cov}{\mathrm{Cov}}
\newcommand{\Corr}{\mathrm{Corr}}
\newcommand{\Prob}{\mathbb{P}}

\newcommand{\F}{\mathcal{F}}
\newcommand{\G}{\mathcal{G}}
\newcommand{\sigB}{\sigma_B^2}
\newcommand{\sigW}{\sigma_W^2}
\newcommand{\neff}{n_{\mathrm{eff}}}

\title{Realized versus Population Reliability:\\
Anytime-Valid Self-Consistency Certification under Correlated Reasoning}

\author{
Terence D. Agbeyegbe\\
Department of Economics\\
Hunter College and Graduate Center, CUNY\\
\texttt{tagbeyeg@hunter.cuny.edu}
\and
Toneukarin G. Agbeyegbe\\
Department of Statistics\\
Rutgers University\\
\texttt{tga17@scarletmail.rutgers.edu}
}

\date{November 2026}

\begin{document}

\maketitle

\begin{abstract}
Self-consistency certification treats repeated sampling from a language model as
sequential evidence accumulation and stops when an anytime-valid e-process crosses
a threshold, guaranteeing finite-sample Type~I error control by Ville's inequality.
Every construction in this line assumes the sampled rollouts are conditionally
independent. We show that this assumption is not innocuous: realistic reasoning
streams are positively correlated. Rollouts share a sampled reasoning prefix, and
aggregate certificates pool a population of prompts of heterogeneous difficulty, and
correlation changes not merely the efficiency of a certificate but \emph{the hypothesis
it certifies}.

We model the dependence through a latent state drawn once per certification unit
(the random-effects / intraclass-correlation structure from clustered data),
inducing exchangeable, equicorrelated increments with correlation $\rho$. Correlation
splits the single null into two: a \emph{realized} null about the drawn belief and a
\emph{population} null about the model's average behavior, which coincide only when
$\rho=0$. Our results are: (i) the standard betting e-process remains anytime-valid,
but for the realized null, not the population null; (ii) read as a population-null
certificate, the asymptotic population-null crossing probability of a suitable mixture
is at least $\pi=\Prob(\mu(Z)>0)$ and need not respect the nominal $\epsilon$;
(iii) a design-effect saturation result: for the within-unit mean, the
variance-equivalent sample size tends to $1/\rho$; additional within-unit
rollouts alone cannot restore population-null validity; and (iv) a prompt-level e-process that restores
population validity, with an allocation rule that, under a pure rollout budget, is
minimized at one rollout per prompt. We confirm the mechanism in simulation. Measuring
it on real rollouts (Claude Haiku~4.5), we find the gap is regime-dependent. On GSM8K,
$\hat\rho=0.435$ but the model is so self-consistent ($\hat\theta=+0.96$) that the gap
is dormant. On AIME the model scatters ($\hat\rho=0.480$, $\hat\theta=-0.15<0$), the
estimated population margin is negative, illustrating the mismatch: read as a population-null
certificate, the single-unit rule produces single-prompt crossings on $24\%$ of prompts,
whereas $\epsilon=0.05$ is a bound for tests of a specified null, while the prompt-level procedure does not certify (empirical
rate $0.00$). The contribution is a
correctness result, not an efficiency one: it identifies which reliability claim an
anytime-valid self-consistency certificate is entitled to make under dependence.
\end{abstract}

%------------------------------------------------------------------
\section{Introduction}
\label{sec:introduction}
%------------------------------------------------------------------

The deployment of large language models (LLMs) in reasoning tasks has revealed a
fundamental tension between reliability and computational cost. Test-time
scaling—including sampling multiple responses, extending context windows, and running additional
reasoning steps—improves output quality but consumes resources. The question of
\emph{when to stop} is both statistical (is there sufficient evidence for the
conclusion?) and economic (is further computation worth the cost?). For
self-consistency \citep{wang2023selfconsistency}, where an LLM is queried repeatedly
at positive temperature and the majority answer is selected, a recent line of work
makes this rigorous: \citet{cordero2025certified} formalize a Martingale Majority
Certificate that, via test supermartingales and Ville's inequality
\citep{ville1939etude}, stops with exact, finite-sample, anytime-valid Type~I error
control. This replaces the heuristic ``sample a fixed number of times'' with a
principled certificate.

\subsection{The gap: dependence}

Every certificate in this line, and the broader anytime-valid toolkit it draws
on \citep{howard2021timeuniform,ramdas2023gametheoretic}, assumes the rollouts are
\emph{conditionally independent} increments with a single drift parameter. While
essential for the underlying theory, this assumption fails to hold in the two regimes
of greatest practical interest.

\paragraph{Shared latent within a prompt.} Tree-of-thought branching, prefix-shared
decoding, and batched or speculative sampling produce rollouts that are independent
only \emph{after} conditioning on a reasoning prefix or plan that is itself sampled
once. Marginally, the rollouts are correlated through the shared prefix.

\paragraph{Across-prompt aggregation.} A deployment-level claim—``this model is
reliable on this task''—pools evidence across a population of prompts. Prompts
differ in difficulty: a model is rock-solid on easy items and scattered on hard ones,
so the per-prompt margin is dispersed. That dispersion is exactly between-unit
variance, and it makes the pooled increments correlated.

In both regimes the increments are positively correlated, and correlation does
something more consequential than inflate variance. It \emph{splits the null
hypothesis in two}. The \emph{realized} null concerns whether, for this prompt's drawn
reasoning, the certified answer is favored; the \emph{population} null asks whether the
answer is favored on average over the model's randomness or the task distribution. When
the increments are independent, these coincide. When they are correlated, they come
apart by exactly the correlation $\rho$, and a procedure that looks valid may be
silently certifying the wrong one.

\subsection{Contributions}

This is a correctness contribution, not an efficiency one. We do not propose a faster
certificate; we identify which reliability claim an anytime-valid certificate is
entitled to make once rollouts are correlated, and we repair it when the population
claim is the target.

\begin{enumerate}[nosep]
  \item \textbf{A dependent evidence model} (Section~\ref{sec:model}). We transport the
  random-effects / intraclass correlation model from clustered data to the reasoning
  stream: a latent state $Z$ is drawn once per certification unit, with rollouts
  conditionally i.i.d.\ given $Z$. This induces exchangeable, equicorrelated increments
  with correlation $\rho = \sigB / (\sigB + \sigW)$, where $\sigB$ and $\sigW$ denote
  the between- and within-unit variances. Consequently, the single null splits into a
  \emph{realized null} ($\mu(Z) \le 0$, governing the unit's conditional mean) and a
  \emph{population null} ($\theta = \E[\mu(Z)] \le 0$, governing the overall mean).

  \item \textbf{Which null is controlled} (Section~\ref{sec:control}). The standard
  betting e-process is \emph{not} a supermartingale under the population null because the
  data eventually reveal $Z$, so the marginal validity argument breaks whenever the
  fraction of units with positive realized drift satisfies $\pi=\Prob(\mu(Z)>0)>0$. But it
  survives, exactly, for the realized null (Theorem~\ref{thm:realized}). Read as a
  population-null certificate, a suitable mixture has asymptotic population-null crossing probability
  at least $\pi$, potentially exceeding the nominal error level $\epsilon$ (Proposition~\ref{prop:overcert}).

  \item \textbf{An impossibility} (Section~\ref{sec:impossibility}). For the within-unit
  sample mean, the variance-equivalent effective sample size tends to $1/\rho$
  (Theorem~\ref{thm:saturation}); the within-unit mean converges to the realized belief $\mu(Z)$,
  and does not in general converge to the population drift $\theta$. Without additional
  structural information or independent units, a realized-null certificate does not yield
  a population-null guarantee.

  \item \textbf{A remedy} (Section~\ref{sec:remedy}). Genuine independence lives
  \emph{across} units, so the e-process must too. A prompt-level e-process on
  within-unit means restores exact population validity
  (Theorem~\ref{thm:promptlevel}), and an allocation analysis shows that, under a pure
  rollout budget, the variance-minimizing choice is \emph{one rollout per prompt}; a
  per-prompt cost is required before extra rollouts help (Proposition~\ref{prop:alloc}).

  \item \textbf{Unknown $\rho$ and beyond exchangeability} (Section~\ref{sec:open}).
  Realized-null validity needs no knowledge of $\rho$; population validity under
  unknown $\rho$ uses variance-adaptive betting. Arbitrary serial dependence within a
  unit is genuinely open and is where dependence-robust (HAC / near-epoch-dependent)
  machinery is required.

  \item \textbf{Measurement} (Section~\ref{sec:experiments}). Simulation confirms the
  mechanism and its repair. On real rollouts (Claude Haiku~4.5) we measure the
  across-prompt $\rho$ on two benchmarks: GSM8K ($\hat\rho=0.435$), where strong
  self-consistency ($\hat\theta=+0.96$) leaves the gap dormant, and AIME
  ($\hat\rho=0.480$, $\hat\theta=-0.15<0$), where the estimated population margin is negative and
  over-certification \emph{occurs}. Read as a population-null test, the single-unit rule
  produces single-prompt crossings on $24\%$ of prompts, versus a nominal $\epsilon=0.05$ population-test level, while the
  prompt-level remedy declines.
\end{enumerate}

\paragraph{Relation to prior certificates.} Our results do not contradict
\citet{cordero2025certified} or its successors. For a single prompt with
independently sampled rollouts $\rho=0$, the two nulls coincide, and those
certificates are exactly correct for the margin they target. The point is that under
the correlation present in shared-prefix decoding and in across-prompt aggregation,
the same construction controls the realized null while the application often wants the
population null. \citet{DeyBhandari2025} study asymptotically optimal sequential
multiple testing in an \emph{equicorrelated} Gaussian model, the same correlation
structure we adopt, but from a fixed-procedure optimality viewpoint; we work in the
anytime-valid e-process framework and isolate the hypothesis-identification problem
that dependence creates. Closest in setting is \citet{ota2026cite}, who give
anytime-valid certification of a target answer as the mode of the response distribution
under data-driven stopping and an unknown answer-category set, with minimax rates and
experiments on open models over hard benchmarks; that work operates under independence and
addresses the unknown category set, whereas we address dependence and the
realized-versus-population identification, which are complementary axes. \citet{akter2026edfl}
similarly build self-normalized empirical-Bernstein e-processes for LLM generation under
unknown centering. Finally, \citet{cordero2025certified} explicitly name applying
anytime-valid certificates to \emph{correlated rollouts} as future work; the present paper
takes up exactly that problem.

\subsection{Outline}

Section~\ref{sec:background} recalls the anytime-valid toolkit as the $\rho=0$
baseline. Section~\ref{sec:model} develops the dependent evidence model and the two
nulls. Section~\ref{sec:control} establishes which null the standard certificate
controls and the over-certification of the population null.
Section~\ref{sec:impossibility} proves the information-saturation impossibility.
Section~\ref{sec:remedy} gives the prompt-level remedy and allocation.
Section~\ref{sec:open} treats unknown $\rho$ and serial dependence.
Section~\ref{sec:experiments} reports simulations and the measured $\rho$.
Section~\ref{sec:discussion} discusses scope and limitations.

%------------------------------------------------------------------
\section{Background: the independent baseline}
\label{sec:background}
%------------------------------------------------------------------

We recall the anytime-valid machinery in the form we need, treating it throughout as
the $\rho = 0$ special case of the dependent model of Section~\ref{sec:model}.

\subsection{Ville's inequality and test supermartingales}

The challenge of sequential testing is \emph{optional stopping}: when the decision to
stop depends on the data, fixed-sample guarantees fail. Ville's inequality is the
remedy.

\begin{definition}[Supermartingale]
\label{def:supermartingale}
An integrable sequence $(M_n)_{n\ge1}$ adapted to a filtration $(\F_n)_{n\ge1}$ is a
\emph{supermartingale} if $\E[M_n\mid\F_{n-1}]\le M_{n-1}$ almost surely for all $n\ge2$.
\end{definition}

\begin{theorem}[Ville, 1939]
\label{thm:ville}
If $(M_n)_{n\ge1}$ is a non-negative supermartingale with $\E[M_1]\le1$, then for any
threshold $u>0$, $\;\Prob\!\left(\sup_{n\ge1}M_n\ge u\right)\le 1/u.$
\end{theorem}

\begin{definition}[Test supermartingale]
\label{def:eprocess}
A \emph{test supermartingale} (the canonical form of an e-process) for a null $H_0$ is a
non-negative supermartingale $(E_n)_{n\ge1}$ under $H_0$ with $\E_0[E_1]\le1$.
\end{definition}

\begin{corollary}[Anytime-valid test]
\label{cor:anytime}
If $(E_n)_{n\ge1}$ is a test supermartingale under $H_0$, the rule
$\tau=\inf\{n\ge1:E_n\ge1/\epsilon\}$ that rejects $H_0$ at $\tau$ satisfies
$\Prob_0(\tau<\infty)\le\epsilon$, regardless of when or whether one stops.
\end{corollary}

\noindent The value $E_n$ is the ``evidence against $H_0$'' accumulated by stage $n$.
Corollary~\ref{cor:anytime} is immediate: $\{\tau<\infty\}=\{\sup_{n\ge1} E_n\ge
1/\epsilon\}$ and Theorem~\ref{thm:ville} applies with $u=1/\epsilon$.

\subsection{Two baseline constructions}

For self-consistency \citep{wang2023selfconsistency}, fix a candidate answer $c$
using a separate pilot sample. One possible increment compares $c$ with a
fixed runner-up $r$:
$Y_n^{\mathrm{pair}}=\mathbf{1}\{X_n=c\}-\mathbf{1}\{X_n=r\}\in\{-1,0,+1\}$,
with population margin $p_c-p_r$. Our experiments instead compare $c$ with
its complement:
$Y_n=\mathbf{1}\{X_n=c\}-\mathbf{1}\{X_n\ne c\}\in\{-1,+1\}$,
with mean $2p_c-1$. Both statistics satisfy $|Y_n|\le B=1$, and the
martingale arguments below apply to either after fixing the statistic.
Unless stated otherwise, $Y_n$, $\mu$ (the conditional mean, formalized in
Section~\ref{sec:model}), and $\theta$ (the population drift) refer to the
leader-versus-complement statistic used in the experiments. The null is
$H_0:\theta\le0$.

\paragraph{Beta-mixture (Martingale Majority Certificate).}
\citet{cordero2025certified} place a Beta prior $\pi$ on the leader probability over the
alternative ($\theta>0$) and form the mixture likelihood-ratio process
$E_n^{\text{Beta}}=\int L_n(\theta)\,\pi(d\theta)$ relative to $\theta=0$, which is a
martingale at $\theta=0$ (and a supermartingale for $\theta\le0$) and crosses $1/\epsilon$
at the boundary $S_n\gtrsim\sqrt{2n(\log(1/\epsilon)+\tfrac12\log n)}$.

\paragraph{Betting / Robbins-stitched.}
For a betting fraction $\lambda\in[0,1/B]$, the product
$M_n=\prod_{t\le n}(1+\lambda Y_t)$ satisfies
$\E_0[1+\lambda Y_t\mid\F_{t-1}]=1+\lambda\theta\le1$ when $\theta\le0$, hence is a
supermartingale; mixing $\lambda$ over a geometric grid
\citep{robbins1970statistical,howard2021timeuniform} yields the law-of-the-iterated-logarithm
boundary $S_n\gtrsim\sqrt{2\sigma^2 n(\log(1/\epsilon)+C\log\log n)}$. The betting form
is the one we carry into the dependent setting, because its supermartingale property
is what dependence disrupts.

\begin{remark}[Scope of this paper]
\label{rem:scope}
The efficiency comparison between these constructions, specifically the $\log n$ versus $\log\log
n$ boundary gap, is orthogonal to our question and we do not pursue it. Both are
valid test supermartingales for $\theta\le0$ \emph{under independence}; our concern is what
``$\theta\le0$'' even means, and which null these processes control, once the
increments are correlated.
\end{remark}

%------------------------------------------------------------------
\section{The Dependent Evidence Model}
\label{sec:model}
%------------------------------------------------------------------

Section~\ref{sec:background} treated the evidence increments as conditionally
independent with a single drift parameter. That is the right model for one
canonical case and the wrong model for several others. When an LLM is queried
multiple times for a fixed prompt with independent samples, the rollouts are
genuinely independent and the constructions of
Section~\ref{sec:background} apply verbatim. But independence fails whenever the
increments share sampled structure: rollouts that branch from a common
reasoning prefix, batched or speculative decoding that reuses randomness, or
an aggregate certificate that pools evidence across a population of prompts to
make a claim about the model rather than about a single query. In each of these
the increments are correlated, and the correlation changes not merely the
efficiency of a certificate but the very hypothesis it certifies.

We formalize this through a latent state drawn once per certification unit and
shared across that unit's increments. The construction is the random-effects
(intraclass-correlation) model familiar from clustered data, transported to the
reasoning stream.

\subsection{Latent-belief evidence processes}
\label{subsec:latent}

\begin{definition}[Latent-belief evidence process]
\label{def:latent}
A \emph{certification unit} emits increments $Y_1,Y_2,\dots$ with cumulative
evidence $S_n=\sum_{t=1}^n Y_t$. A \emph{latent state} $Z$, taking values in a
measurable space $\mathcal{Z}$, is drawn once at the start of the unit.
Conditional on $Z$, the increments $(Y_t)_{t\ge1}$ are independent and
identically distributed with
\[
  \mu(Z):=\E[Y_t\mid Z],
  \qquad
  \Var(Y_t\mid Z)=\sigma_W^2(Z),
  \qquad
  |Y_t|\le B \text{ a.s.}
\]
\end{definition}

We collect the marginal moments that govern the process:
\[
  \theta:=\E[\mu(Z)],
  \qquad
  \sigB:=\Var\!\big(\mu(Z)\big),
  \qquad
  \sigW:=\E\!\big[\sigma_W^2(Z)\big],
  \qquad
  \sigma^2:=\sigB+\sigW .
\]
Here $\theta$ is the \emph{population drift}, $\sigB$ the
\emph{between-unit} variance (dispersion of the belief), $\sigW$ the mean
\emph{within-unit} variance, and $\sigma^2=\Var(Y_t)$ the marginal
variance by the law of total variance.

Two filtrations matter, and the gap between them drives every result that
follows. The \emph{observed} filtration is
$\F_n=\sigma(Y_1,\dots,Y_n)$; it is what a certificate actually sees.
The \emph{oracle} filtration $\G_n=\sigma(Z,Y_1,\dots,Y_n)$
additionally resolves the latent state. Crucially, $Z\notin\F_n$: the
decision-maker never observes the belief, only its noisy consequences.

\begin{proposition}[Exchangeability and equicorrelation]
\label{prop:equicorr}
Under Definition~\ref{def:latent} with $\sigW > 0$, the increments $(Y_t)$
are exchangeable, and for every $t\ne s$,
\[
  \Cov(Y_t,Y_s)=\sigB,
  \qquad
  \Corr(Y_t,Y_s)=\rho:=\frac{\sigB}{\sigB+\sigW} \in [0,1).
\]
The common pairwise correlation $\rho$ is exact, irrespective of any
heterogeneity in the conditional variances $\sigma_W^2(Z)$.
\end{proposition}

\begin{proof}
Exchangeability is immediate from conditional i.i.d.\ structure. For $t\ne s$,
the law of total covariance and conditional independence give
\[
  \Cov(Y_t,Y_s)
  =\E\!\big[\Cov(Y_t,Y_s\mid Z)\big]
   +\Cov\!\big(\E[Y_t\mid Z],\E[Y_s\mid Z]\big)
  =0+\Var\!\big(\mu(Z)\big)=\sigB .
\]
Dividing by the marginal variance $\sigma^2=\sigB+\sigW > 0$ yields
$\rho$.
\end{proof}

\begin{proposition}[Variance of the within-unit mean]
\label{prop:neff}
Let $\bar{Y}_n=S_n/n$. Under Definition~\ref{def:latent} with $\sigW > 0$,
\[
  \Var(\bar{Y}_n)=\sigB+\frac{\sigW}{n}
  =\frac{\sigma^2}{\neff},
  \qquad
  \neff:=\frac{n}{1+(n-1)\rho}\;\xrightarrow[n\to\infty]{}\;\frac{1}{\rho},
\]
with the convention $1/0 := \infty$ when $\rho=0$.
\end{proposition}

\begin{proof}
By Proposition~\ref{prop:equicorr},
$\Var(\bar{Y}_n)=n^{-2}\big[n\sigma^2+n(n-1)\sigB\big]
=\sigW/n+\sigB$. Substituting $\sigB=\rho\sigma^2$ and
$\sigW=(1-\rho)\sigma^2$ gives
$\Var(\bar{Y}_n)=\sigma^2\{1+(n-1)\rho\}/n$, hence the stated
$\neff$.
\end{proof}

Proposition~\ref{prop:neff} is the seed of the impossibility result in
Section~\ref{sec:impossibility}: when $\rho>0$, averaging within a single unit
cannot drive the variance below $\sigB$, so the within-unit mean is a
consistent estimator of the realized belief $\mu(Z)$, not of the population
drift $\theta$.

\begin{remark}[Recovery of the independent model]
\label{rem:rho0}
$\rho=0$ iff $\sigB=0$ iff $\mu(Z)=\theta$ almost surely. In that case the
latent state is degenerate, $\F_n$ and $\G_n$ carry the same
information about the drift, and Definition~\ref{def:latent} collapses to the
conditionally independent process of Section~\ref{sec:background}. The
constructions and guarantees recalled there are the $\rho=0$ special case of
what follows.
\end{remark}

\begin{remark}[Why the observed filtration misbehaves]
\label{rem:filtration}
Under the oracle filtration the conditional mean is constant,
$\E[Y_{n+1}\mid\G_n]=\mu(Z)$. Under the observed filtration it is
the posterior mean of the belief given the data,
$\E[Y_{n+1}\mid\F_n]=\E[\mu(Z)\mid\F_n]$, which is
not constant in $n$ and converges to $\mu(Z)$ as the data accumulate. The
observed process therefore \emph{learns} the realized drift over time. This is
the mechanism that breaks the marginal supermartingale property under the
population null and forces the distinction between the two hypotheses below;
we make it precise in Section~\ref{sec:control}.
\end{remark}

\subsection{Two null hypotheses}
\label{subsec:twonulls}

Correlation splits the single null of Section~\ref{sec:background} into two
hypotheses that coincide only in the degenerate case.

\begin{definition}[Population and realized nulls]
\label{def:twonulls}
The \emph{population null} concerns the model's average behavior,
\[
  H_0^{\mathrm{pop}}:\ \theta\le 0
  \qquad\text{versus}\qquad
  H_1^{\mathrm{pop}}:\ \theta>0 .
\]
The \emph{realized null} concerns the drawn belief,
\[
  H_0^{\mathrm{rel}}:\ \mu(Z)\le 0
  \qquad\text{versus}\qquad
  H_1^{\mathrm{rel}}:\ \mu(Z)>0 .
\]
\end{definition}

\begin{remark}[Equivalence only at $\rho=0$]
\label{rem:coincide}
Since $\mu(Z)=\theta$ a.s.\ iff $\sigB=0$ iff $\rho=0$, the two nulls are
equivalent precisely when the increments are uncorrelated. Whenever $\rho>0$
they are distinct events, and a procedure that controls one need not control
the other.
\end{remark}

The two nulls answer different questions. The realized null asks whether, for
\emph{this} unit's drawn reasoning, the certified outcome is favored: a claim
about the particular prompt-and-prefix in front of the decision-maker. The
population null asks whether the outcome is favored on average over the model's
randomness or over the population of prompts: a claim about the system. For a
single high-stakes query the realized null may be exactly what one wants; for a
deployment-level reliability guarantee the population null is the target. The
two come apart by $\rho$, and conflating them is the error this paper is about.

\begin{definition}[Realized-alternative probability]
\label{def:pi}
Write $\pi:=\Prob\!\big(\mu(Z)>0\big)$ for the probability that the realized
belief favors the certified outcome.
\end{definition}

Under $H_0^{\mathrm{pop}}$ with $\sigB>0$ one may still have $\pi>0$: the
average belief disfavors the outcome while individual draws favor it. This
wedge between $\theta\le0$ and $\pi>0$ is the quantity that a single-unit
certificate silently exploits, and it reappears as the over-certification rate
in Section~\ref{sec:control}.

\subsection{Instantiation: dependent self-consistency}
\label{subsec:sc-dep}

We recast the self-consistency process of Section~\ref{sec:background} inside
the latent-belief model. Let $c$ be the answer selected on an independent pilot
sample and fix it before the held-out evidence stream begins. In the empirical
implementation $Y_t=2\mathbf{1}\{X_t=c\}-1\in\{-1,+1\}$, with $B=1$.
Its conditional mean is $\mu(Z)=2p_c(Z)-1$; a positive value means that
$c$ receives more than half the conditional answer probability, not merely
that it is the most probable individual answer. The latent state admits two
readings, both of practical relevance.

\emph{Across-prompt aggregation.} Let $Z$ index the prompt together with its
terminal answer distribution, drawn from the task distribution; conditional on
$Z$ the held-out rollouts are independent, with $\mu(Z)=2p_c(Z)-1$ the
leader-versus-complement margin for that prompt. The population drift
$\theta=\E[2p_c(Z)-1]$, incorporating the random pilot-selection
procedure, is the average margin over the task. Thus $H_0^{\mathrm{pop}}$
states that a pilot-selected answer has probability at most one half on average. Here $\rho$ measures how much the leader margin varies
across prompts relative to within-prompt sampling noise.

\emph{Shared latent within a prompt.} Let $Z$ be a sampled reasoning prefix or
latent plan that a batch of rollouts conditions on, as in tree-of-thought or
prefix-shared decoding; conditional on $Z$ the continuations are independent,
but marginally they are correlated through the shared prefix. Even a
single-prompt certificate then accumulates evidence about $\mu(Z)$, the
margin under the drawn prefix, rather than about the prompt's unconditional
margin.

\begin{remark}[Relation to independent self-consistency]
\label{rem:mmc}
For a single prompt with independently sampled rollouts there is no shared
latent, $\rho=0$, and the certificate controls the corresponding
fixed-candidate margin ($2p_c-1$ in our empirical implementation) exactly; the guarantees of Section~\ref{sec:background} are
unaffected. The dependent regime arises only under the two structures above. We
state this explicitly because the distinction is easy to miss: the issue is not
that anytime-valid self-consistency certificates are invalid, but that under
correlation they certify the realized null when the population null is what the
application requires.
\end{remark}

\begin{remark}[The estimand under a pilot-selected leader]
\label{rem:estimand}
When the answer $c$ is fixed from a separate pilot slice of rollouts (as in
Section~\ref{sec:experiments}), the held-out score is
$Y_t=2\mathbf{1}\{X_t=c\}-1$. The realized mean $\mu(Z)$ measures the
conditional majority probability of the pilot-selected answer, not its
probability of being correct or its margin over the second most likely answer.
The population mean $\theta$ averages this score across freshly drawn prompts
\emph{and} their corresponding random pilots. Fixing $c$ before observing the
held-out sequence preserves the conditional martingale argument. For the
population-level construction, independent units must include independently
drawn pilots; conditioning on all realized pilots is not generally compatible
with the common-mean assertion $\E[U_q]=\theta$.
\end{remark}

\subsection{Assumptions}
\label{subsec:assumptions}

We restate the regularity conditions of Section~\ref{sec:background}
conditionally on the latent state.

\begin{assumption}[Conditional sub-Gaussian increments]
\label{ass:subgauss-cond}
There exists $\sigma_W>0$ such that for all $\lambda\in\R$ and $t\ge1$,
\[
  \E\!\left[e^{\lambda(Y_t-\mu(Z))}\mid Z\right]
  \le e^{\lambda^2\sigW/2}\quad\text{a.s.}
\]
\end{assumption}

\begin{remark}
Assumption~\ref{ass:subgauss-cond} holds for bounded increments with
$\sigW=B^2$ by Hoeffding's lemma applied conditionally on $Z$. It governs
concentration \emph{within} a unit; the between-unit dispersion $\sigB$ is
unrestricted beyond having a finite second moment.
\end{remark}

\begin{assumption}[Exchangeable dependence]
\label{ass:exch}
The increments follow the latent-belief structure of
Definition~\ref{def:latent}: conditionally i.i.d.\ given a single latent state
$Z$ drawn once per unit.
\end{assumption}

Assumption~\ref{ass:exch} is the working dependence model for the rest of the
paper. It captures shared-belief correlation exactly but excludes arbitrary
serial dependence within a unit, where the conditioning argument of
Section~\ref{sec:control} no longer applies. We return to that more general
case, and to the tools required for it, in Section~\ref{sec:open}.

%------------------------------------------------------------------
\section{What the Certificate Controls}
\label{sec:control}
%------------------------------------------------------------------

We now apply the standard betting e-process to a correlated stream and determine
which hypothesis it certifies. The headline is a dichotomy: the marginal
supermartingale property fails under the population null but the certificate remains
exactly valid for the realized null, and reading it as a population-null certificate
can invalidate population-null Type~I error control. Table~\ref{tab:scope} summarizes the
scope of each result, its target null, the filtration on which it holds, and the
dependence assumption, so the central distinction is not misread; none of the validity
guarantees requires knowledge of $\rho$.

\begin{table}[h]
\centering
\caption{Scope of the main results. ``Anytime'' denotes exact, finite-sample validity
at data-dependent stopping times. None of the validity guarantees requires knowing
$\rho$; $\rho$ governs only efficiency (betting fraction and allocation).}
\label{tab:scope}
\small
\resizebox{\textwidth}{!}{%
\begin{tabular}{lllll}
\toprule
Result & Target & Filtration & Dependence & Type \\
\midrule
Thm~\ref{thm:realized} (realized validity)      & $H_0^{\mathrm{rel}}:\mu(Z)\le0$  & oracle $\G_n$ (cond.\ on $Z$) & exchangeable          & finite-sample, anytime \\
Prop~\ref{prop:overcert} (over-certification)   & $H_0^{\mathrm{pop}}:\theta\le0$  & observed $\F_n$               & exch., $\pi>0$         & asymptotic lower bound \\
Thm~\ref{thm:saturation} (saturation)           & estimation of $\theta$           & observed $\F_n$               & exch., $\rho>0$        & exact variance, a.s.\ limit \\
Thm~\ref{thm:promptlevel} (remedy)              & $H_0^{\mathrm{pop}}:\theta\le0$  & unit-level $\mathcal{H}_q$     & indep.\ across units   & finite-sample, anytime \\
\bottomrule
\end{tabular}}
\end{table}

\subsection{The marginal supermartingale property fails}

Consider the betting e-process $M_n=\prod_{t=1}^n(1+\lambda Y_t)$ with
$\lambda\in(0,1/B]$ applied to the observed stream. Under the population null
$\theta\le0$ one might hope $M_n$ is an $\F_n$-supermartingale, as in
Section~\ref{sec:background}. It is not.

\begin{proposition}[Failure under the population null]
\label{prop:fail}
Under Definition~\ref{def:latent}, suppose $\theta\le0$ but $\pi=\Prob(\mu(Z)>0)>0$.
Then $M_n=\prod_{t\le n}(1+\lambda Y_t)$ with $\lambda\in(0,1/B]$ is \emph{not} an
$\F_n$-supermartingale under $H_0^{\mathrm{pop}}$. If instead $\mu(Z)\le0$ almost
surely (so $\pi=0$), the conditioning argument of Theorem~\ref{thm:realized} shows
$M_n$ \emph{is} a supermartingale and no failure occurs; thus $\pi>0$ is the operative
condition, not $\sigB>0$ alone.
\end{proposition}

\begin{proof}
By Remark~\ref{rem:filtration}, $\E[Y_{n+1}\mid\F_n]=\E[\mu(Z)\mid\F_n]$, the
posterior mean of $\mu(Z)$ given the data. Hence
$\E[M_{n+1}\mid\F_n]=M_n\,(1+\lambda\,\E[\mu(Z)\mid\F_n])$. On the event that the
observed history favors a positive belief, which has positive probability whenever
$\pi>0$, the posterior mean $\E[\mu(Z)\mid\F_n]$ is positive even though
$\theta\le0$, so $\E[M_{n+1}\mid\F_n]>M_n$ there. The supermartingale inequality is
violated, so Ville's inequality cannot be invoked under $H_0^{\mathrm{pop}}$ on the
observed filtration.
\end{proof}

The break is not pathological: it is the process \emph{learning} the realized belief.
The repair is to condition on $Z$.

\subsection{Validity for the realized null}

\begin{theorem}[Realized-null validity]
\label{thm:realized}
Let $(E_n)$ be the betting e-process $M_n=\prod_{t\le n}(1+\lambda Y_t)$ with
$\lambda\in[0,1/B]$, or its stitched mixture over a fixed grid of such $\lambda$, built
from the increments $(Y_t)$, and $\tau=\inf\{n:E_n\ge1/\epsilon\}$. Under
Definition~\ref{def:latent},
\[
  \Prob\!\big(\{\textstyle\sup_n E_n\ge1/\epsilon\}\cap\{\mu(Z)\le0\}\big)\le\epsilon,
  \qquad\text{and}\qquad
  \Prob\!\big(\text{certify}\mid\mu(Z)\le0\big)\le\epsilon .
\]
That is, the procedure is an exact, anytime-valid test of the \emph{realized} null
$H_0^{\mathrm{rel}}:\mu(Z)\le0$.
\end{theorem}

\begin{proof}
Work on the oracle filtration $\G_n=\sigma(Z,Y_1,\dots,Y_n)$. Conditional on
$Z=z$ with $\mu(z)\le0$, the increments are i.i.d.\ with non-positive mean, so for the
betting process $\E[1+\lambda Y_t\mid\G_{t-1},Z=z]=1+\lambda\mu(z)\le1$ and $(E_n)$ is
a non-negative $\G_n$-supermartingale with $\E[E_1\mid Z=z]\le1$ under
$\Prob(\cdot\mid Z=z)$. (For the stitched mixture the same holds by
convexity: it is an average over a fixed grid of $\lambda$ of conditional
supermartingales, and a convex combination of conditional supermartingales is one.) Ville's inequality applied conditionally gives
$\Prob(\sup_n E_n\ge1/\epsilon\mid Z=z)\le\epsilon$ for a.e.\ such $z$. Averaging the
conditional bound over $\{\mu(Z)\le0\}$ yields
$\Prob(\text{certify}\mid\mu(Z)\le0)\le\epsilon$, and multiplying by
$\Prob(\mu(Z)\le0)\le1$ gives the joint bound.
\end{proof}

\begin{remark}[On the Beta-mixture]
\label{rem:beta-cond}
We state Theorem~\ref{thm:realized} for betting-based e-processes because their
supermartingale property holds conditionally on $Z$ by the one-line argument above.
The Beta-mixture certificate of \citet{cordero2025certified} is also realized-null
valid, but this requires specifying its likelihood under the latent model: the Beta
prior must be placed on the \emph{conditional} leader probability $p_c(Z)$, so that the
mixture is a test supermartingale for the conditional law given $Z$. Under that reading
the same conditioning argument applies; we restrict the statement to the betting form
to avoid asserting it for an unspecified likelihood.
\end{remark}

\begin{remark}[What this buys and does not]
\label{rem:realized-reading}
Theorem~\ref{thm:realized} is good news for single-query certification: even under
shared-prefix correlation, the standard certificate exactly controls the error of the
claim ``for the reasoning actually drawn, this answer is favored.'' It needs no
knowledge of $\rho$. What it does \emph{not} deliver is a statement about the model's
average behavior, and the next result shows the two can be far apart.
\end{remark}

\subsection{Over-certification of the population null}

\begin{proposition}[Over-certification]
\label{prop:overcert}
Suppose $H_0^{\mathrm{pop}}$ holds ($\theta\le0$) but
$\pi=\Prob(\mu(Z)>0)>0$. For a fixed $0<\lambda<1/B$ let
$g_\lambda(z)=\E[\log(1+\lambda Y_t)\mid Z=z]$ and
$A_\lambda=\{z:g_\lambda(z)>0\}$. The betting process
$M_n(\lambda)=\prod_{t\le n}(1+\lambda Y_t)$ then satisfies
\[
 \lim_{n\to\infty}\Prob\bigl(\sup_{t\le n}M_t(\lambda)
             \ge1/\epsilon\bigr)\ge\Prob(Z\in A_\lambda).
\]
In particular, let $\lambda_j\in(0,1/B)$ decrease to zero and let
$w_j>0$ with $\sum_jw_j=1$. The mixture
$E_n=\sum_jw_jM_n(\lambda_j)$ satisfies
\[
 \lim_{n\to\infty}\Prob\bigl(\sup_{t\le n}E_t
             \ge1/\epsilon\bigr)\ge\pi.
\]
These crossing probabilities refer to the population-null data-generating
law and need not be bounded by $\epsilon$. The second conclusion does not
necessarily hold for a fixed finite betting grid.
\end{proposition}

\begin{proof}
Conditional on $Z=z$, bounded increments are i.i.d. For each fixed
$0<\lambda<1/B$, the strong law gives
$n^{-1}\log M_n(\lambda)\to g_\lambda(z)$ almost surely. Hence
$M_n(\lambda)\to\infty$ on $A_\lambda$, establishing the first claim.
If $\mu(z)>0$, then $g_0(z)=0$ and
$\frac{d}{d\lambda}g_\lambda(z)|_{\lambda=0}=\mu(z)>0$.
Thus some sufficiently small positive $\lambda_j$ has
$g_{\lambda_j}(z)>0$. The positive-weight term
$w_jM_n(\lambda_j)$ then diverges, and so does $E_n$.
Integrating this conditional almost-sure statement over
$\{\mu(Z)>0\}$ and using continuity from below gives the second bound.
\end{proof}

\begin{remark}[Lower bound, not equality]
\label{rem:lb}
The lower bound $\pi$ applies to the countable mixture specified in
Proposition~\ref{prop:overcert}, not necessarily to the finite grid used in the
experiments. On realized-negative units ($\mu(Z)\le0$), conditional
supermartingale validity bounds additional crossing probability by at most
$\epsilon(1-\pi)$. At finite sample sizes, positive-drift units may not
have crossed even if their asymptotic log-growth is positive. The qualitative point
is what matters: the error is governed by $\pi$, a quantity unrelated to the nominal
level $\epsilon$.
\end{remark}

\begin{remark}[The practical reading]
\label{rem:practical-overcert}
Draw many rollouts from one prompt (or pool a difficult task) and certify the
plurality answer: a threshold crossing supplies evidence about the realized conditional
margin (subject to its test error), not the average margin over prompts. With positively correlated
rollouts these are different claims, and the gap depends on the distribution of realized means, including $\pi$;
$\rho$ alone does not determine $\pi$.
A certificate calibrated to $\epsilon=0.05$ may have a population-null
crossing rate much larger than $\epsilon$, as the proposition demonstrates.
\end{remark}

%------------------------------------------------------------------
\section{The Impossibility: Information Saturation}
\label{sec:impossibility}
%------------------------------------------------------------------

Over-certification might seem fixable by sampling harder, that is, by taking more rollouts from the same
unit. It is not. Under $\rho>0$ a single unit carries a bounded amount of information
about the population drift, no matter the budget.

\begin{theorem}[Information saturation]
\label{thm:saturation}
Under Definition~\ref{def:latent} with $\rho>0$:
\begin{enumerate}[nosep]
  \item $\bar Y_n\to\mu(Z)$ almost surely, and $\Var(\bar Y_n)\to\sigB>0$; the
  within-unit mean is consistent for the realized belief $\mu(Z)$ and is
  \emph{not} a consistent estimator of $\theta$.
  \item The variance-equivalent effective sample size for the within-unit sample
  mean is $\neff=n/(1+(n-1)\rho)\to 1/\rho$. This is a design-effect
  calculation for that estimator, not a universal information bound for
  every estimator or structured parametric model.
\end{enumerate}
\end{theorem}

\begin{proof}
Conditional on $Z$, the increments are i.i.d.\ with mean $\mu(Z)$, so the strong law
gives $\bar Y_n\to\mu(Z)$ a.s.; the variance statement and $\neff$ are
Proposition~\ref{prop:neff}. Since $\mu(Z)$ is non-degenerate when $\sigB>0$,
$\bar Y_n$ does not converge to the constant $\theta$.
\end{proof}

\begin{corollary}[Single-unit budget guarantees are vacuous for the population null]
\label{cor:vacuous}
The high-probability budget guarantee of the independent theory—``with probability
$1-\delta$ the certificate stops within $O(\sigma^2\log(1/\epsilon)/\theta^2)$
rollouts''—bounds the time to resolve the \emph{realized} drift $\mu(Z)$. As a
statement about the population drift $\theta$ it is vacuous as $n$ grows: additional samples from the same unit reduce
$\Var(\bar Y_n)$ negligibly and add essentially no information about $\theta$.
\end{corollary}

\begin{remark}[The design effect]
\label{rem:designeffect}
Theorem~\ref{thm:saturation} is the cluster-sampling \emph{design effect} \citep{kish1965}
$1+(n-1)\rho$ transplanted to rollouts. It is the sharp negative result the
independent theory lacks: it explains why the cheap move of repeatedly sampling one prompt
cannot deliver a population guarantee, and points to what must replace it.
\end{remark}

%------------------------------------------------------------------
\section{The Remedy: Certifying Across Units}
\label{sec:remedy}
%------------------------------------------------------------------

If $\theta$ is the target, the genuine independence lives \emph{across} units, so the
e-process must be built there.

\subsection{A prompt-level e-process}

\begin{definition}[Prompt-level evidence]
\label{def:promptlevel}
Index independent certification units $q=1,2,\dots$ (e.g.\ distinct prompts drawn from
the task). For unit $q$ draw $n_q$ rollouts and form the within-unit mean
$U_q:=\bar Y_{n_q}^{(q)}\in[-B,B]$. The $U_q$ are independent across $q$ with
$\E[U_q]=\theta$ and $\Var(U_q)=\sigB+\sigW/n_q$.
\end{definition}

\begin{theorem}[Population validity]
\label{thm:promptlevel}
For a betting fraction $\lambda\in[0,1/B]$, the prompt-level process
$M_Q:=\prod_{q=1}^Q(1+\lambda U_q)$ is a non-negative supermartingale under
$H_0^{\mathrm{pop}}:\theta\le0$ with $\E_0[M_1]\le1$. Hence
$\tau=\inf\{Q:M_Q\ge1/\epsilon\}$ satisfies $\Prob_0(\tau<\infty)\le\epsilon$ as an
exact, anytime-valid test of the population null. Mixing $\lambda$ over a geometric
grid recovers the law-of-the-iterated-logarithm boundary in the number of units $Q$.
\end{theorem}

\begin{proof}
Let $\mathcal{H}_q=\sigma(U_1,\dots,U_q)$. The $U_q$ are independent with mean
$\theta$, so $\E[1+\lambda U_q\mid\mathcal{H}_{q-1}]=1+\lambda\theta\le1$ when
$\theta\le0$; the supermartingale property and $\E_0[M_1]\le1$ follow, and
Corollary~\ref{cor:anytime} gives the bound. The stitched mixture over $\lambda$
applies unchanged at the unit level.
\end{proof}

The key shift is the index: evidence is accumulated over \emph{units}, each
contributing a single near-independent summary $U_q$, rather than over correlated
rollouts within one unit.

\subsection{Budget allocation}

How should a fixed budget be split between more units and more rollouts per unit? The
answer inverts the naive intuition.

\begin{proposition}[Allocation]
\label{prop:alloc}
Suppose every unit draws $n$ rollouts and there are $Q$ units, for a total of
$N=Qn$ rollouts, with $\sigB>0$. The variance of the pooled estimator
$\hat\theta=Q^{-1}\sum_q U_q$ is
\[
  \Var(\hat\theta)=\frac{\sigB+\sigW/n}{Q}
  =\frac{n\sigB+\sigW}{N},
\]
which is strictly increasing in $n$. Under a pure rollout budget $N$, the
variance-minimizing allocation is therefore $n=1$: one rollout per unit, spread over
as many units as possible. If instead each unit carries a fixed prefill/encoding cost
$c_p>0$ and each rollout a cost $c_r>0$, with total cost $C=Q(c_p+nc_r)$ held fixed, the
optimum is
\[
  n^\star_{\mathrm{continuous}}=\sqrt{\frac{c_p}{c_r}\cdot\frac{\sigW}{\sigB}}
        =\sqrt{\frac{c_p}{c_r}\cdot\frac{1-\rho}{\rho}} .
\]
The feasible minimizer over integer $n\ge1$ is obtained by comparing the
admissible integers nearest $\max\{1,n^\star_{\mathrm{continuous}}\}$.
\end{proposition}

\begin{proof}
The first display is $\Var(U_q)/Q$ with $\Var(U_q)=\sigB+\sigW/n$ and
$Q=N/n$; monotonicity in $n$ is immediate. For the two-cost model, substitute
$Q=C/(c_p+nc_r)$ into $\Var(\hat\theta)=(\sigB+\sigW/n)/Q$ to obtain
$\Var(\hat\theta)=C^{-1}(\sigB+\sigW/n)(c_p+nc_r)$; differentiating in $n$
and setting the derivative to zero gives $c_r\sigB=c_p\sigW/n^2$, i.e.\
$n^\star_{\mathrm{continuous}}=\sqrt{(c_p/c_r)(\sigW/\sigB)}$, and
$\sigW/\sigB=(1-\rho)/\rho$ by Proposition~\ref{prop:equicorr}.
\end{proof}

\begin{remark}[Correcting the naive intuition]
\label{rem:alloc}
The instinct that more rollouts per prompt always helps is wrong for the population
null: under a pure rollout budget it strictly \emph{hurts}, because rollouts past the
first are correlated and contribute less than independent units. Extra rollouts per
unit pay off only to amortize a genuine per-unit cost, and even then only up to
$n^\star_{\mathrm{continuous}}\propto\sqrt{(1-\rho)/\rho}$, which shrinks as correlation grows.
\end{remark}

%------------------------------------------------------------------
\section{Unknown $\rho$ and Beyond Exchangeability}
\label{sec:open}
%------------------------------------------------------------------

\paragraph{Unknown $\rho$.} Realized-null validity (Theorem~\ref{thm:realized}) needs
no knowledge of $\rho$; it follows purely from exchangeability. Population validity
(Theorem~\ref{thm:promptlevel}) also holds for any fixed $\lambda$ regardless of
$\rho$, since it conditions only on independence \emph{across} units. What depends on
$\rho$ is efficiency: the optimal betting fraction and allocation involve
$\Var(U_q)=\sigB+\sigW/n$, which is unknown when $\rho$ is. We therefore
recommend a variance-adaptive bet at the unit level, such as an empirical-Bernstein betting
fraction in the style of \citet{waudbysmith2024estimating}, or the Catoni
construction of \citet{wang2023catoni} for heavy-tailed summaries, which is valid
without a variance bound and adapts $\lambda$ to the realized spread of the $U_q$.
This keeps the population guarantee exact while letting the procedure learn the
correlation it faces.

\paragraph{Beyond exchangeability.} The conditioning argument of
Section~\ref{sec:control} uses Assumption~\ref{ass:exch}: a single latent state drawn
once per unit, with conditional independence given $Z$. It does not cover arbitrary
serial dependence \emph{within} a unit. When $Y_t$ depends on $Y_{t-1}$ in an
unrestricted way, there is no $Z$ to condition on and the supermartingale property is
not recovered by enlarging the filtration. Handling that regime requires either a
mixing or near-epoch-dependence rate on the within-unit process, or a block
construction that forms approximately independent summaries from contiguous blocks of
rollouts. This is the natural point of contact with dependence-robust inference
(HAC/near-epoch-dependent estimation), and we flag it as the principal open direction
rather than claim it solved.

%------------------------------------------------------------------
\section{Experiments}
\label{sec:experiments}
%------------------------------------------------------------------

We report three studies. A controlled simulation confirms every theoretical claim and
exhibits the over-certification mechanism and its repair (Section~\ref{subsec:synthetic}).
We then measure the across-prompt $\rho$ on two real benchmarks chosen to straddle the
boundary: GSM8K, where a strong model is nearly always self-consistent, and AIME, where it
is not (Section~\ref{subsec:real}), and on AIME exhibit the over-certification
\emph{occurring} and the prompt-level remedy correcting it
(Section~\ref{subsec:overcert}). All runs fix \texttt{SEED}~$=456$; difficulty is read off a
pilot slice of rollouts and every estimate is computed on a disjoint held-out slice, so
selecting on difficulty cannot bias $\theta$, $\pi$, or $\rho$.

\subsection{Synthetic validation of the mechanism}
\label{subsec:synthetic}

We generate latent-belief rollouts: a per-unit belief $\mu(Z)$ is drawn from a
mixture calibrated to $\rho=0.09$, and conditional on it the increments are sampled
i.i.d. We estimate $\rho$ by the one-way ANOVA intraclass correlation, track
$\Var(\bar Y_n)$ and $\neff$, and run both the single-unit certificate and the
prompt-level remedy.

\begin{table}[h]
\centering
\caption{Synthetic study (population null true, $\theta\le0$, $\pi=0.37$,
planted $\rho=0.09$). The single-unit certificate controls the realized null but
over-certifies the population null toward $\pi$; the prompt-level remedy restores
population validity, including at the least-favorable $\theta=0$.}
\label{tab:synthetic}
\begin{tabular}{lcc}
\toprule
Quantity & Single-unit certificate & Prompt-level remedy \\
\midrule
Recovered $\rho$ (ICC) & \multicolumn{2}{c}{$0.091$ (planted $0.091$)} \\
$\Var(\bar Y_n)$ limit & \multicolumn{2}{c}{$0.090=\sigB$ (saturates)} \\
$\neff$ limit & \multicolumn{2}{c}{$\approx 11 \approx 1/\rho$} \\
\midrule
Realized-null Type~I & $0.005$ & N/A \\
Population-null certification rate & $0.32\ (\to\pi=0.37)$ & $0.033$ at $\theta=0$ \\
\bottomrule
\end{tabular}
\end{table}

The numbers track the theory exactly: the ICC recovers the planted $\rho$
(Proposition~\ref{prop:equicorr}); $\Var(\bar Y_n)$ flattens onto $\sigB$ and
$\neff$ caps near $1/\rho$ (Theorem~\ref{thm:saturation}); the single-unit certificate
holds the realized null far below $\epsilon=0.05$ while certifying the population null
at $0.32$ and climbing toward $\pi=0.37$ even though $\theta\le0$
(Proposition~\ref{prop:overcert}); and the prompt-level construction restores
validity, sitting at $0.033\le0.05$ at the binding $\theta=0$
(Theorem~\ref{thm:promptlevel}). The conservativeness ($<\epsilon$) is expected of
anytime-valid tests and is not under-powering. Under a pure rollout budget the variance is minimized at one
rollout per unit, and the $\sqrt{(c_p/c_r)(1-\rho)/\rho}$ rule appears only under the
two-cost model (Proposition~\ref{prop:alloc}).

\subsection{Measuring $\rho$ on two real benchmarks}
\label{subsec:real}

The synthetic study sets $\rho$ by hand. Two empirical questions determine whether the theoretical gap has practical consequences: whether $\rho>0$ on real models, and, separately, whether the population null ($\theta\le0$) ever holds alongside $\pi>0$.

 We measure the across-prompt $\rho$ on two benchmarks
with a single production model (Claude Haiku~4.5) at temperature~$1$: GSM8K
\citep{cobbe2021gsm8k} and AIME \citep{aime19832024}. For each prompt a pilot slice fixes
the leader and the prompt's difficulty; the disjoint held-out rollouts are scored $\pm1$
against that fixed leader; and $\rho$ is the one-way ICC on the held-out matrix with a
cluster-bootstrap interval (resample prompts, recompute the ICC, take the $2.5/97.5$
percentiles). The split removes the data-adaptive-leader bias of scoring against a leader
chosen from the same rollouts.

\begin{table}[h]
\centering
\caption{Across-prompt correlation and the boundary, two benchmarks (Claude Haiku~4.5,
temperature~$1$, split-sample leader). On both, $\hat\rho$ is bounded well away from zero.
The benchmarks differ in $\theta$: GSM8K is easy and the leader is strongly favored on
average, so the realized--population gap is structurally present but operationally
invisible; AIME is hard and the leader is \emph{not} favored on average, placing the
negative estimated population margin and positive estimated $\pi$, the regime where over-certification can occur.}
\label{tab:realrho}
\begin{tabular}{lccccc}
\toprule
Benchmark (prompts $\times$ rollouts) & $\hat\rho$ (95\% CI) & $1/\hat\rho$
 & $\hat\theta$ & $\hat\pi$ & median agr. \\
\midrule
GSM8K \;($300\times30$)  & $0.435\ [0.208,0.619]$ & $2.3$ & $+0.96$ & $0.99$ & $1.00$ \\
AIME \;($150\times30$)   & $0.480\ [0.405,0.544]$          & $2.1$ & $-0.15$ & $0.33$ & $0.30$ \\
\bottomrule
\end{tabular}

\smallskip
{\footnotesize Both intervals are cluster bootstraps over prompts, $B=2000$
(Remark~\ref{rem:bootstrap}).}
\end{table}

Two readings. First, the premise holds on both benchmarks: $\hat\rho=0.44$ on GSM8K and
$0.48$ on AIME, with effective per-prompt sample sizes of only about two. By
Theorem~\ref{thm:saturation} this is a budgeting fact independent of difficulty, since the diminishing-returns effect depends on the intraclass correlation and
does not establish a hard cutoff after three rollouts.
Second, and this is what GSM8K alone could not show, the two benchmarks sit on opposite
sides of the boundary. On GSM8K $\hat\theta=+0.96$: the model is so self-consistent that
realized and population conclusions agree for essentially every prompt, and the gap of
Proposition~\ref{prop:overcert}, though present ($\rho>0$), has nothing to act on. This
holds in aggregate across difficulty: $283$ of $300$ prompts are unanimous and only $17$
fall below pilot agreement $0.9$ (too few to certify either way), with held-out plurality
accuracy $0.97$, so no GSM8K regime with appreciable mass approaches the boundary, and in
contrast to AIME, where binning by pilot agreement sweeps $\theta$ from $-0.70$ to $+0.97$
(Table~\ref{tab:buckets}). On AIME
the model scatters; its pilot-selected answer receives less than half of held-out answers on average, so
$\hat\theta=-0.15<0$ with $\hat\pi=0.33$, indicating an estimated negative population drift, while about one-third of
prompts have positive observed held-out margins. The held-out plurality matches the ground-truth answer
in only $38\%$ of AIME prompts, consistent with $\theta<0$ reflecting genuinely low
reliability rather than an artifact of leader selection. (Accuracy and the self-consistency
margin $\theta$ are related but distinct: a model can be confidently consistent on a wrong
answer, so $40\%$ accuracy bounds difficulty, not $\theta$ directly.)

\begin{remark}[Inference caveat]
\label{rem:bootstrap}
On easy prompts the $\pm1$ scores pile at the boundary (GSM8K median agreement $1.00$),
violating the within-group normality the classical F-based interval assumes. We therefore
report \emph{cluster bootstraps} for both benchmarks: resample prompts with replacement,
recompute the ICC, take the $2.5/97.5$ percentiles, which serves as the defensible interval for
clustered, non-normal scores.
\end{remark}

\begin{remark}[Interpretation of empirical margins]
\label{rem:margins}
The reported $\hat\theta$ averages held-out scores over sampled prompts and
pilot-selected leaders. The fraction $\hat\pi$ classifies units by the sign of
their finite-sample held-out mean, and is therefore a descriptive proxy for
$\Prob(\mu(Z)>0)$ rather than an error-free estimate of latent signs. In
particular, a percentage of promptwise crossings is not an estimate of the
Type~I error rate for a population test: repeated independent realizations
under a specified population-null law would be needed to estimate that rate.
The prompt-bootstrap crossing frequency is likewise a descriptive resampling
statistic, not a proof of the nominal size bound.
\end{remark}

\subsection{Over-certification, exhibited and repaired (AIME)}
\label{subsec:overcert}

On AIME we run the single-unit betting certificate on each prompt's held-out stream and the
prompt-level certificate across prompts, at $\epsilon=0.05$.

\begin{table}[h]
\centering
\caption{Over-certification on AIME (Claude Haiku~4.5, $150$ prompts). At the
whole-benchmark level the estimated population margin is negative ($\hat\theta=-0.15$), yet single-unit
certificates fire on nearly one prompt in four, while the prompt-level remedy declines.}
\label{tab:overcert}
\begin{tabular}{lc}
\toprule
Quantity (whole benchmark; $\hat\theta=-0.15$, $\hat\pi=0.33$) & Rate \\
\midrule
Single-unit certification rate (per prompt) & $0.240$ \\
Prompt-level population certification rate (bootstrap) & $0.000$ \\
Nominal level $\epsilon$ & $0.05$ \\
\bottomrule
\end{tabular}
\end{table}

\begin{table}[h]
\centering
\caption{AIME by difficulty (pilot-agreement bins; estimates on held-out). As difficulty
falls (left to right), $\theta$ sweeps from $-0.70$ to $+0.97$ while $\pi$ and the
single-unit certification rate track the \emph{realized} side. The drift crosses zero
between the $(0.5,0.7]$ and $(0.7,0.9]$ bins, where the single-unit rate jumps from $0.06$
to $0.73$.}
\label{tab:buckets}
\begin{tabular}{lccccc}
\toprule
Pilot-agreement bin & $n$ & $\hat\theta$ & $\hat\pi$ & $\hat\rho$ & single-unit rate \\
\midrule
$[0.0,0.3]$ & $60$ & $-0.70$ & $0.03$ & $0.11$ & $0.00$ \\
$(0.3,0.5]$ & $37$ & $-0.34$ & $0.16$ & $0.14$ & $0.03$ \\
$(0.5,0.7]$ & $16$ & $-0.14$ & $0.25$ & $0.11$ & $0.06$ \\
$(0.7,0.9]$ & $11$ & $+0.74$ & $1.00$ & $0.07$ & $0.73$ \\
$(0.9,1.0]$ & $26$ & $+0.97$ & $1.00$ & $0.04$ & $1.00$ \\
\bottomrule
\end{tabular}
\end{table}

\begin{figure}[h]
\centering
\includegraphics[width=0.82\linewidth]{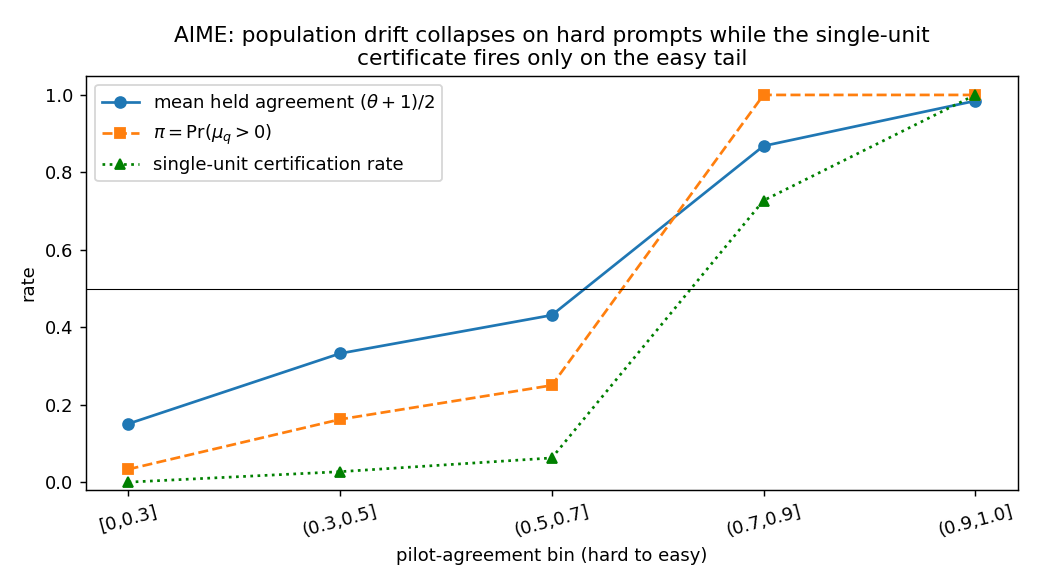}
\caption{The empirical wedge on AIME. Across pilot-agreement bins (hard at left, easy at
right): mean held agreement $(\theta+1)/2$ and $\pi=\Prob(\mu_q>0)$ rise together, while the
single-unit certification rate stays near zero across the low-agreement bins, which hold
most of the mass and pull the population drift to $\hat\theta=-0.15$, and climbs to $1.00$
only on the high-agreement minority. That minority is the $24\%$ the single-unit certificate
fires on, even though the estimated population margin is negative.}
\label{fig:wedge}
\end{figure}

The contrast in Table~\ref{tab:overcert} is the central empirical result. The estimated population margin is negative on AIME ($\hat\theta=-0.15$): the model's pilot-selected answer is not,
on average, favored in held-out rollouts. Yet, used as a test of the population null, the
single-unit certificate produces crossings on $24.0\%$ of individual prompts, exceeding the nominal
$\epsilon=0.05$ population-test level numerically, while the prompt-level certificate, which accumulates one near-independent
summary per prompt, declines in these prompt-bootstrap resamples (observed rate $0.000$). Table~\ref{tab:buckets} and Figure~\ref{fig:wedge} show the
mechanism: the single-unit crossings come entirely from the high-agreement prompts, on which
the model is confidently consistent, while the large mass of scattered hard prompts pulls the
population drift negative. The drift crosses zero between the middle bins, where the
single-unit rate jumps from $0.06$ to $0.73$.

\paragraph{Reading the contrast correctly.} The $24.0\%$ of single-unit certifications are
\emph{not} errors in the realized sense. By Theorem~\ref{thm:realized} each crossing is covered by a test whose error probability is controlled
under its realized null; no individual crossing proves that its alternative is true; crossings on such prompts may reflect positive realized margins, but their
latent signs cannot be determined exactly from the finite held-out samples. The data bear this out directly:
among the $101$ prompts with nonpositive \emph{estimated held-out means} not one single-unit
certificate fires ($0/101$), a descriptive observation consistent with realized-null validity, and the entire $24.0\%$ comes from the $49$ prompts with positive estimated held-out means (firing
rate $0.74$), which should not be equated with proof of true positive conditional means. The over-certification is relative to the
\emph{population} null: reading those certifications as evidence that the model is reliable on
AIME is the error, because the estimated average leader-versus-complement margin is negative. A
single-prompt certificate answers a narrower question than a deployment-level reliability
claim, and under $\rho=0.48$ the two diverge by enough that the aggregate single-unit
certification rate ($0.24$) bears no relation to the nominal $\epsilon$. The prompt-level
construction targets the population null directly; its lack of observed crossings
is consistent with, but does not establish, the validity theorem.

\paragraph{Regime dependence, empirically.} GSM8K and AIME together make the point that
neither makes alone: the realized--population gap is one phenomenon with two faces. On an easy
task with a strong model it is structurally present ($\rho=0.44$) but dormant, because
$\theta$ is far from the boundary; on a hard task it is empirically salient ($\hat\rho=0.48$, $\hat\theta<0$),
with promptwise crossings on $24\%$ of sampled prompts despite a negative
mean held-out margin. The distinction
matters exactly where reliability is genuinely in question.

\paragraph{Scope of the measurement.} We measure the \emph{across-prompt} $\rho$; the
within-prompt shared-prefix $\rho$, which threatens even single-prompt certificates, requires
tree-of-thought or prefix-shared generation and is not measured here. The reported runs use a
single named production model; a reproducible open-model replication (e.g.\ on
\texttt{gpt-oss-20b} or \texttt{Qwen3-30B}) is the natural robustness check but could alter the estimated magnitudes and even the qualitative pattern.

\subsection{Protocol and reproducibility}
\label{subsec:protocol}

All runs fix \texttt{SEED}$=456$ and query Claude Haiku~4.5 (snapshot
\texttt{claude-haiku-4-5-20251001}) via the Anthropic Messages API at temperature~$1.0$
with $8$ concurrent workers. \emph{GSM8K}: the public test split
\citep{cobbe2021gsm8k}, $300$ prompts $\times\,30$ rollouts, \texttt{max\_tokens}$=512$,
prompt ``\emph{Solve the problem. Show brief reasoning, then end with a line `The answer
is $\langle$number$\rangle$'.}''; a pilot of $10$ rollouts fixes the leader and the
disjoint $20$ held-out rollouts are scored. \emph{AIME}: the 1983--2024 set
\citep{aime19832024} (HuggingFace \texttt{datasets-server}, no token required), $150$
integer-answer problems $\times\,30$ rollouts, \texttt{max\_tokens}$=1024$, prompt
``\emph{Solve this AIME problem. The answer is an integer from 0 to 999\ldots}''; a
pilot of $10$ rollouts fixes the leader and difficulty bin and the disjoint $20$
held-out rollouts are scored $\pm1$. Answers are extracted by a regular expression
preferring an explicit ``answer is'' or \texttt{\textbackslash boxed\{\}} field and
otherwise the trailing integer. Both the single-unit and prompt-level certificates use
the stitched betting mixture
$E=\frac{1}{|\Lambda|}\sum_{\lambda\in\Lambda}\prod_{t}(1+\lambda Y_t)$ over the grid
$\Lambda=\{0.1,0.2,0.3,0.4,0.5\}$, rejecting when $E\ge1/\epsilon=20$
($\epsilon=0.05$); $\rho$ is the one-way $\mathrm{ICC}(1)$ with a cluster bootstrap over
prompts ($B=2000$). The source code and saved rollout matrices accompany this submission as 
supplementary material for peer review, and will be deposited in a public repository upon publication.

%------------------------------------------------------------------
\section{Discussion}
\label{sec:discussion}
%------------------------------------------------------------------

\subsection{What the guarantees do}

\paragraph{They identify the hypothesis.} The central service of this paper is not a
new certificate but a clarification: under correlation, an anytime-valid
self-consistency certificate controls the realized null
(Theorem~\ref{thm:realized}), an exact and useful guarantee for single-query
decisions, and silently differs from the population null by $\rho$. Naming which
claim is being made is the prerequisite for trusting it.

\paragraph{They remain anytime-valid.} Both the realized-null reading and the
prompt-level population construction are exact and finite-sample, holding at
data-dependent stopping times. The repair changes the index of accumulation, not the
underlying Ville machinery.

\paragraph{They are honest about budget.} The effective-sample-size cap $1/\rho$
converts a vague intuition into a number: with $\hat\rho\approx0.45$, a single prompt
is worth only about two independent rollouts for the population question, no matter how
many one draws.

\subsection{What they do not}

\paragraph{No guarantee under arbitrary serial dependence.} The exchangeable
latent-belief model is the working assumption; unrestricted within-unit serial
dependence is open (Section~\ref{sec:open}).

\paragraph{No detection guarantee.} As in the independent theory, Type~I control is
not power: near the boundary $\theta\approx0$, certification of the population null may
require many units.

\paragraph{Practical impact is regime-dependent} On easy benchmarks with strong models the
realized--population gap, though structurally present ($\rho>0$), is operationally
small because $\theta$ is large; the distinction matters most where reliability is
genuinely in question.

\subsection{Limitations and future work}

The real-data study measures the across-prompt $\rho$; the within-prompt
shared-prefix $\rho$, which threatens even single-prompt certificates, requires
tree-of-thought or prefix-shared generation and is not measured here. The over-certification
is demonstrated on AIME, where the estimated population margin is negative ($\hat\theta<0$) and
single-unit certificates fire on $24\%$ of the observed prompts while the prompt-level procedure does not cross;
a reproducible open-model replication is the natural robustness check. A hierarchical
(prefix-tree) correlation model would refine Theorem~\ref{thm:saturation}'s constant for
branching rollouts, and the serial-dependence extension via mixing/near-epoch-dependence
rates is the principal theoretical opening. We regard the present contribution as resolving the identification question of \emph{which} reliability an anytime-valid certificate certifies under correlation, isolating the conditions under which over-certification materializes, and providing the construction that restores the population guarantee when that is the target.

%------------------------------------------------------------------
\bibliographystyle{plainnat}

\end{document}